\documentclass[pra]{revtex4}
\usepackage{graphicx}
\usepackage{amsmath}
\usepackage{amssymb}

\begin{document}
\title{Optimization of electron pumping by harmonic mixing}
\author{Niklas Rohling, Frank Grossmann}
\affiliation{Institut f\"ur Theoretische Physik, Technische Universit\"at
Dresden, D-01062 Dresden, Germany}
\date{March 9, 2011}

\begin{abstract}
For a symmetric bridge coupled to infinite leads, in the 
presence of a dipole-coupled external ac-field with harmonic mixing,
we solve the Schr\"odinger equation in the time-domain using open boundary
conditions as well as in the energy-domain using Floquet scattering
theory.
As this potential breaks parity and generalized parity, we find a
non-vanishing average current. We then optimize the relative amplitude
ratio between the fundamental and the second harmonic leading to a
maximum in the pump current.
\end{abstract}
\maketitle

\section{Introduction}
Generating non-vanishing average electronic currents in statically 
unbiased systems, i.e. electron pumping, by external ac-fields has been 
realized in experiments with quantum dots \cite{Swetal99,DiCM03,VDiCM05},
nanotubes \cite{Letal05}, semiconductor heterostructures 
\cite{Letal99,Letal02} and a Josephson junction array \cite{Metal03}.

Theoretically, the adiabatic case of slow driving has been treated amongst
others by Thouless \cite{thouless}, by Brouwer \cite{brouwer}, who used a
scattering approach, and by Zhou \textit{et al.} \cite{fzhou}, applying
Keldysh-Green's function methodology. The non-adiabatic,
time-periodic case can be handled in the Floquet formalism,
formally treating the system as time-independent.
An early example of a similar strategy is the heuristic approach to
understand the effect of time-periodic driving on the current
voltage characteristics of superconductor-insulator-superconductor
junctions that has been given by Tien and Gordon \cite{TG63}.
More recently, Floquet scattering theory has been used to get expressions 
for transmission probabilities \cite{BL82,LR99,tang,LR00,martinez,zhou,emmanouilidou}. 
In addition, Kim \cite{kim} as well as Moskalets and B\"uttiker \cite{mb} 
exploited this method for considering an electronic pump consisting of two
oscillating $\delta$-peaks. In \cite{BL82,LR99,tang,LR00,zhou,kim,mb,
emmanouilidou} the scattering matrix  is determined from the matching
conditions of the wave function. Furthermore, electron-pumping scenarios have
been treated with the help of Floquet theory using an equation-of-motion
approach in the Heisenberg  picture \cite{camalet,kohler},  solving the
master equation \cite{lehmann,goychuk,kohler}, or using non-equilibrium
Green's functions \cite{stafford,arrachea1,arrachea2}. The authors of
\cite{camalet,kohler,lehmann,goychuk,arrachea1} have considered tight-binding
Hamiltonians, while in \cite{stafford} an interacting two-level system has
been studied. \cite{arrachea2} provides a comparison of Green's function
theory with the scattering matrix.

Recently, Kurth and coworkers \cite{kurth} proposed a time-dependent 
approach for quantum transport, treating the leads by using open boundaries
for the central region. This method is not restricted to 
time-periodic problems.  In the same Reference a simple scheme for
ac-transport is studied as one application. Stefanucci \textit{et al.}
\cite{stefanucci} later used this algorithm of time-evolution for the
investigation of systems where a net current is generated by a traveling
wave in the potential of  the Schr\"odinger equation.

In this paper, we consider a harmonic-mixing
dipole field 
as another potential that shows the pumping effect. 
In order to break (generalized)
parity, the external field is composed of a fundamental
frequency component together with an additional second harmonic term.
Harmonic mixing has been studied previously in the tight-binding case
\cite{kohler,goychuk} as well as purely classically \cite{marchesoni}.
In Section \ref{sec:floquet}, in order to calculate the transport across a
structure under the influence of the external ac-field, we first review
Floquet scattering theory, allowing us to arrive at an expression for the
{\it stationary} net current from the matching conditions via the 
scattering matrix.

Secondly, in Section \ref{sec:time}, we treat the transport problem from
a {\it time-dependent} point of view, focusing on the transient dynamics of
the current after a sudden switching on of the driving.  To this end we are
considering the zero temperature case and are choosing an equilibrium state
of the undriven system as initial condition \cite{cini}.
We review the algorithm of time-evolution from Ref. \cite{kurth}, where an
expression for the time-dependent current is gained by using a Schr\"odinger
equation with open boundaries.
In Section \ref{sec:dipole}  we explicitly apply both ways to calculate
the current in the problem of electron pumping and it 
is shown that the time-average of the current over one period converges
in the long-time limit to the result of Floquet theory. That was to be 
expected as there
are no bound states in the considered system which could lead to a
non-convergent time dependence as shown in the work of Khosravi
\textit{et al.} \cite{khosravi}. Our studies extend previous work in the
monochromatic case by Li and Reichl \cite{LR00} and lead to
optimal values for the relative amplitudes of the first and
second harmonic as a function of the incoming energy.
In Section \ref{sec:concs} we give conclusions and an outlook.

\section{Floquet scattering theory}
\label{sec:floquet}

Floquet scattering theory has been employed to calculate transmission
probabilities in tunneling systems driven by a monochromatic laser 
field \cite{LR99,HDR01}. Here we briefly review the formalism 
in order to apply it to the phenomenon of electron pumping.

We consider the time-dependent Schr\"odinger equation (TDSE) with one 
dimension in space (atomic units (a.u.) are used throughout the paper),
\begin{equation}
i\partial_t\Psi(x,t)=\left[-\frac{1}{2}\partial_x^2+V(x,t)\right]\Psi(x,t).
\end{equation}
The potential shall fulfill the following conditions:
\begin{enumerate}
 \item $V(x,t+T)=V(x,t)$ (periodicity in time),
\item there are $x_L<x_R\in \mathbb{R}$ with
$V(x,t)=0~\text{if}~x<x_L~\text{or if}~x>x_R$,
\item within $x_L<x<x_R$ we know the analytical solution of the TDSE,
\item the potential is bounded.
\end{enumerate}
We denote the left lead ($x<x_L$) with $L$, the right one ($x>x_R$) with $R$,
and the central region ($x_L<x<x_R$) with $C$. Due to the time periodicity,
i.e.\ condition 1 above, solutions of the 
TDSE can be written in the form \cite{kohler}
\begin{eqnarray}
\label{eqn:solution}
 \Psi(x,t)&=&e^{-i\epsilon t} \phi(x,t),
\\
\phi(x,t+T)&=&\phi(x,t)=\sum_{n=-\infty}^\infty c_n(x)e^{-in\omega t},
\end{eqnarray}
where $\epsilon$ denotes the so called Floquet energy or quasi-energy,
the frequency $\omega$ is given by $\omega=2\pi/T$.

The second condition for $V(x,t)$ leads to the following 
form of the solution in the region $x<x_L$,
\begin{equation}
\label{eqn:links}
 \Psi(x,t)=e^{-i\epsilon t}
\sum_{n=-\infty}^{\infty}\left(\frac{a_L^n}{\sqrt{k_n}}e^{ik_nx}+
\frac{b_L^n}{\sqrt{k_n}}e^{-ik_nx}\right)e^{-in\omega t}
\end{equation}
and for $x>x_R$,
\begin{equation}
\label{eqn:rechts}
 \Psi(x,t)=e^{-i\epsilon t}
\sum_{n=-\infty}^{\infty}\left(\frac{a_R^n}{\sqrt{k_n}}e^{-ik_nx}
+\frac{b_R^n}{\sqrt{k_n}}e^{ik_nx}\right)e^{-in\omega t}
\end{equation}
with the wavenumbers
\begin{equation}
 k_n=\sqrt{2(\epsilon+n\omega)}.
\end{equation}
The Floquet scattering matrix $S$ connects the outgoing
current amplitudes $b_\alpha^n$ with the ingoing
ones $a_\alpha^n$:
\begin{equation}
\label{eqn:fl-s-matrix}
b_\alpha^n=\sum_{\beta=L,R}\sum_mS_{\alpha\beta}^{nm}a_\beta^m.
\end{equation}
To quantify the net current $\langle I\rangle = 1/T \int_0^T{\rm d}t~I(t)$,
we have to include a summation over the Floquet modes in the Landauer
formula \cite{mb} as an incoming wave with energy $E$ can be scattered to an
outgoing one with energy $E+n\omega$ ($n\in\mathbb{Z}$). In the zero
temperature case and including a factor two for the spin this leads to
\begin{equation}
\label{eqn:strom}
 \langle I\rangle = 
\frac{1}{\pi}\int\limits_0^{E_F} {\rm d}E
\sum_{m\geq m_0}\left[|S_{RL}^{m0}(E)|^2-|S_{LR}^{m0}(E)|^2\right]
\equiv \int\limits_0^{E_F} {\rm d}E~\frac{{\rm d}\langle I\rangle}{{\rm d}E}.
\end{equation}
Here the sum is only taken over those $m$ which fulfill $m\geq m_0$ with
$m_0=-E/\omega$. That means bound states, for which $k_n$ is imaginary, 
do not contribute to the current.

$|S_{RL}^{mn}(E)|^2=|S_{LR}^{mn}(E)|^2$ holds and thus $\langle I\rangle$
vanishes if the potential fulfills parity ($V(-x,t)=V(x,t)$) \cite{mb} or
generalized parity ($V(-x, t+T/2)=V(x,t)$) \cite{emmanouilidou}.
As we are interested in non-vanishing net current the potential
to be studied below is supposed to break (generalized) parity. This can be
achieved either by a breaking of the symmetry in position space 
(ratchet effect) and with an additional monochromatic field
or by temporal symmetry breaking (harmonic mixing) 
\cite{kohler}. We will consider the second case. In order to calculate the
relevant part of the Floquet scattering matrix that appears in 
Eq.\ (\ref{eqn:strom}), we use matching conditions for the solutions
and their derivatives at the boundaries of region $C$. More details for
the case to be considered in Sec. \ref{sec:dipole} can be
found in Appendix A.

The formalism can be generalized for cases in which the analytical solution
is not known for the whole region $C$ but for sections composing $C$ and
for cases in which the potential in the leads is not zero but
position-independent.

\section{Time evolution}
\label{sec:time}
Discretization of the TDSE in space (interval length $\Delta x$) leads to
a tridiagonal matrix for the Hamiltonian,
\begin{equation}
 H=\left(\begin{array}{ccccc}\ddots & \ddots &  &  & \\
          \ddots& h_{i-1} & n & 0 & \\
           & n & h_i & n & \\
           & 0 & n & h_{i+1} & \ddots\\
           & & & \ddots & \ddots
         \end{array}
\right)
\end{equation}
with
\begin{equation}
 h_i=\frac{1}{(\Delta x)^2}+V(x_i,t)\hspace{1cm}\text{and}\hspace{1cm}n
=-\frac{1}{2(\Delta x)^2}.
\end{equation}
Note that $H=H(t)$ is explicit time-dependent 21
By splitting $\psi$ into the projections on the three regions $x<x_L$,
$x\in[x_L,x_R]$ and $x>x_R$ and calling $\psi$ in the respective region
$\psi_L$, $\psi_C$ and $\psi_R$ the TDSE takes a form which has been
studied by Hellums and Frensley \cite{HF94},
\begin{equation}
\label{eqn:form}
i\partial_t\left(\begin{array}{c}\psi_L\\\psi_C\\\psi_R\end{array}\right)
= \left(\begin{array}{ccc}H_{LL} & H_{LC} & 0 \\
                          H_{CL} & H_{CC}(t) & H_{CR} \\
                               0 & H_{RC} & H_{RR}\end{array}
  \right)
\left(\begin{array}{c}\psi_L\\
                      \psi_C\\
                      \psi_R\end{array}
\right).
\end{equation}
$H_{\alpha\alpha}$ ($\alpha=L,C,R$) are tridiagonal matrices and 
$H_{C\alpha}$, $H_{\alpha C}$ ($\alpha=L,R$) have only one non-zero entry.
We restrict our discussion in the remainder of this paper to the
case of an explicit time dependence of the Hamiltonian
only in the central region.

The idea of handling (\ref{eqn:form}) is to treat only the central part
$x\in[x_L,x_R]$ explicitly as an open quantum system and get the influence
of the leads (that are regions $x<x_L$ and $x>x_R$) by a source-term giving
the influence of the wave function in the leads and a memory-term giving
the feedback of the part of $\psi_C(t=0)$ which propagates into the leads.
This has been done in \cite{HF94} for a static Hamiltonian by calculating
expressions for the propagator in the leads.

Kurth \textit{et al.} \cite{kurth} presented a numerical scheme for solving
equations given in the form (\ref{eqn:form})
even for cases with a time- but not position-dependent potential in the leads
using a generalized Cayley method. In the case of zero potential in the leads,
it has the form
\begin{equation}
\label{eqn:cayley}
(1+i\delta H^{(m)})\psi^{(m+1)} = (1-i\delta H^{(m)})\psi^{(m)},
\end{equation}
wherein $(m)$ denotes the index of time,
$H^{(m)}=\frac{1}{2}\left(H(t_{m+1})+H(t_m)\right)$, $t_m=m\Delta t$ and
$\delta=\Delta t/2$ is a half time step in the time discretization.
Splitting (\ref{eqn:cayley}) into parts according to the spatial
segmentation and employing the parts for $L$ and $R$ in the one 
for $C$ leads to
\begin{equation}
 [1+i\delta H_{\rm eff}^{(m)}]\psi_C^{(m+1)} = 
[1-i\delta H_{\rm eff}^{(m)}]\psi_C^{(m)}+\sum_{\alpha=L,R}T_\alpha^{(m)}
\end{equation}
with $H_{\rm eff}^{(m)}$ and $T_\alpha^{(m)}$ defined as
\begin{equation}
 H_{\rm eff}^{(m)}:=\left(H^{(m)}_{CC}
  -i\delta H_{CL}(1+i\delta H_{LL})^{-1}H_{LC}
  -i\delta H_{RC}(1+i\delta H_{RR})^{-1}H_{RC}\right),
\end{equation}
\begin{equation}
\label{eqn:sou_mem}
 T_\alpha^{(m)}:=-i\delta H_{C\alpha} \left(1+
 \frac{1-i\delta H_{\alpha\alpha}}{1+i\delta H_{\alpha\alpha}}\right)
 \psi_{\alpha}^{(m)}.
\end{equation}
As $\psi_{\alpha}^{(m)}$ will not be calculated, it has to be replaced in
(\ref{eqn:sou_mem}) via the projection of (\ref{eqn:cayley}) onto $L$,
respectively $R$. The result reads
\begin{equation}
\label{eqn:memusour}
\begin{split}
 T_\alpha^{(m)}= & -\delta^2 H_{C\alpha} \sum_{k=1}^m\left[
\frac{(1-i\delta H_{\alpha\alpha})^{m-k}}{(1+i\delta H_{\alpha\alpha})^{m-k+1}}
+\frac{(1-i\delta H_{\alpha\alpha})^{m-k+1}}{(1+i\delta H_{\alpha\alpha})^{m-k+2}}
\right]H_{\alpha C}(\psi_C^{(k)}+\psi_C^{(k-1)})\\
& -2i\delta H_{C\alpha}
\frac{(1-i\delta H_{\alpha\alpha})^m}{(1+i\delta H_{\alpha\alpha})^{m+1}}
\psi_\alpha^{(0)}.
\end{split}
\end{equation}
The first line of (\ref{eqn:memusour}) is called memory term as it describes
the influence of the history of $\psi_C$ up to the time $t_m$. The second line
gives the effect of the initial wave function in the leads and is called source
term. In \cite{kurth} a detailed instruction for implementing the equations
above is given. In this work we follow those lines.

At $t=0$ we start in the ground state of the total system without the 
time-dependent part of the potential, which is switched on at $t=0$. 
For the calculation of the current, we have to take into account all 
values of the wavenumber $k$ between zero and $k_{F}=\sqrt{2E_{F}}$
where $E_{F}$ denotes the Fermi energy. For each $k$ there are two linear
independent wave functions. In the leads they are given by
Eq.\ (\ref{eqn:links}, \ref{eqn:rechts}) with $a^n_L=0$,
$a^n_R=\delta_{0n}$ (denoted as $\psi^R_k$)
and with $a^n_L=\delta_{0n}$, $a^n_R=0$ (denoted as $\psi^L_k$).
The $b$-coefficients follow from the (standard) scattering matrix for $t<0$.
The initial wave function in the central region are obtained from the
matching conditions in a similar way as the scattering matrix.
We get the current by integration over $k$ \cite{kurth}:
\begin{equation}
\label{eqn:strom_za}
 I(x,t) = \int\limits_0^{k_{F}}{\rm d}k~ \underbrace{\frac{1}{\pi}
\Im\left((\psi_k^L)^*\partial_x\psi_k^L 
+ (\psi_k^R)^*\partial_x\psi_k^R\right)}_{{\rm d}I/{\rm d}k}
\end{equation}
To compare this to the calculation with the Floquet scattering matrix 
we use that ${\rm d}I/{\rm d}E={\rm d}I/{\rm d}k\times1/k$.

As it is done in \cite{stefanucci} we define the temporal mean
value of the current $\langle I\rangle(x,t)$ (for fixed position $x$) by
\begin{equation}
\label{eqn:<I>(t)}
 \langle I\rangle (x,t):=\Theta(T-t)\frac{1}{t}\int_0^t {\rm d}t'~I(x,t')
 +\Theta(t-T)\frac{1}{T}\int_{t-T}^t{\rm d}t'~I(x,t').
\end{equation}
As a typical behavior of an open system without bound states we expect that 
deviation from the solution of Floquet scattering vanishes for
$t\to\infty$ and thus $\lim_{t\to\infty}\langle I\rangle(x,t)=\langle I\rangle$
with $\langle I\rangle$ from (\ref{eqn:strom}). Furthermore, we stress
that the time-dependent formalism with open boundaries goes beyond
standard wave-packet approaches 
\cite{henseler2,frank1,frank,lefebvre,jaeaeskelaeinen,gaerttner,kramer},
yielding only transmission probabilities and not allowing for a 
description of the transient time-evolution of  the composite system dynamics.

\section{Harmonic-mixing dipole field}
\label{sec:dipole}

In the remainder of this paper we study the transport induced by 
a dipole field in the central region of our system. The
potential in the leads is assumed to be zero and the generalized 
parity is broken by harmonic mixing according to either
\begin{equation}
\label{eqn:dipole1}
 V(x,t)=\Theta(t)\Theta(d/2-|x|) x \left(A\sin(\omega t)+B\cos(2\omega t)\right)
\end{equation}
or
\begin{equation}
\label{eqn:dipole2}
 V(x,t)=\Theta(t)\Theta(d/2-|x|) x \left(A\cos(\omega t)+B\sin(2\omega t)\right).
\end{equation}
We refer to the potentials (\ref{eqn:dipole1}), (\ref{eqn:dipole2}) as 
case I, respectively II, and will concentrate on the optimization of
the relative strengths $A$ and $B$ of the first and second harmonic later-on.
 
The analytical solution of the TDSE within $|x|<d/2$ and a scheme for 
extracting the Floquet scattering matrix $S$ using the matching conditions 
is given in Appendix \ref{sec:Anhang A} for (\ref{eqn:dipole1}). Case II of
Eq. (\ref{eqn:dipole2}) can be treated accordingly.\\
\begin{figure}[hbtp]
\begin{center}
\includegraphics[clip]{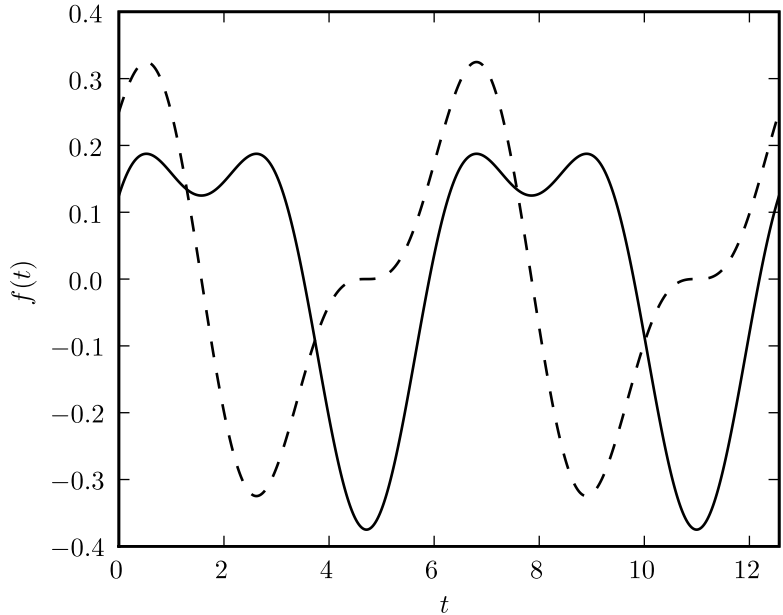}
\caption{$f(t)=\frac{1}{4}\left[\sin(t)+\frac{1}{2}\cos(2t)\right]$ (solid) and
$f(t)=\frac{1}{4}\left[\cos(t)+\frac{1}{2}\sin(2t)\right]$ (dashed line)}
\label{fig:sin_cos}
\end{center}
\end{figure}

As can be seen in Figure \ref{fig:sin_cos} there is a qualitative difference 
in the potentials described by (\ref{eqn:dipole1}) and (\ref{eqn:dipole2}). 
Especially time-reversal {\it parity} ($V(-x,-t)=V(x,t)$) is present in 
(\ref{eqn:dipole2}) after a time shift of $\pi/(2\omega)$ while it is broken 
in (\ref{eqn:dipole1}), where time-reversal {\it symmetry} 
($V(x,-t)=V(x,t)$) is fulfilled after $t\rightarrow t+\pi/(2\omega)$. 

\subsection{Asymptotic average current}

\begin{figure}[hbtp]
\begin{center}
\includegraphics[clip]{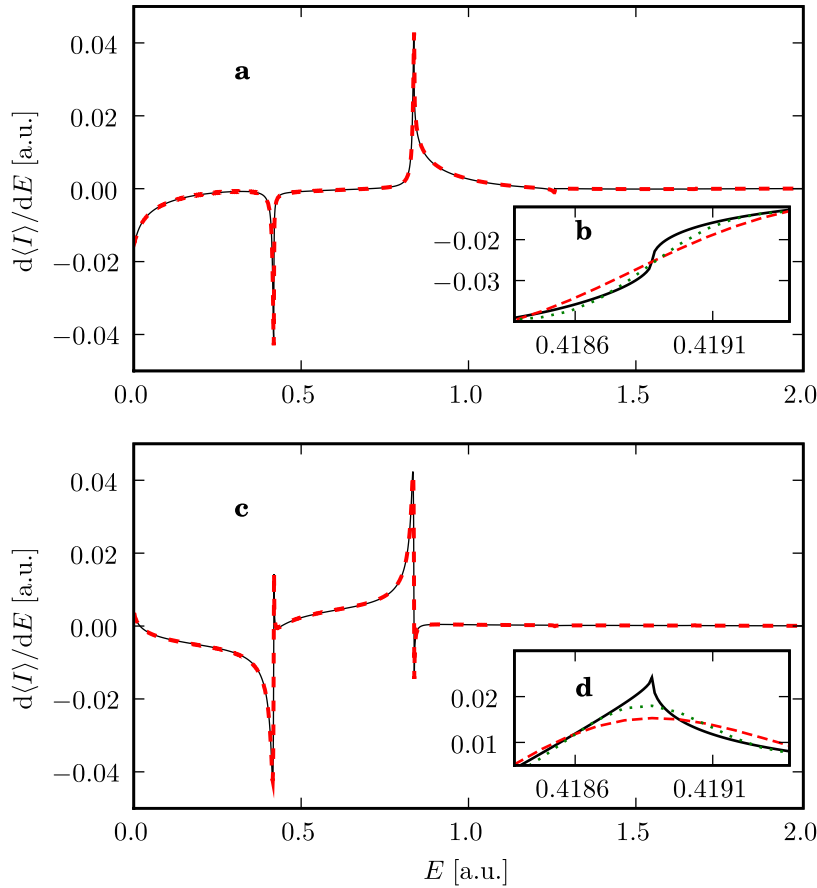}
\caption{(Color online) ${\rm d}\langle I\rangle/{\rm d}E$ for case I 
(a, b) and case II (c, d) calculated with Floquet scattering theory 
(black solid lines) and time-dependent solution (red dashed lines). Parameters
are in both cases $d=3$~a.u., $A=2B=0.25$~a.u.,
$\omega=2\pi/15~\mathrm{a.u.}\approx0.419~\mathrm{a.u.}$ Numerical parameters
of time evolution are $\Delta x=0.01$~a.u. and $\delta=0.05$~a.u.
The time-dependent result is plotted for $t=5000$~a.u. and $x=0$
(the green dotted lines in the insets show the result for $t=10000$~a.u.).
The scattering matrix has been computed with $11$  Floquet modes for
(a), (c) and $17$ for the insets.}
\label{fig:dipole}
\end{center}
\end{figure}

Although cases I and II have different symmetry properties, our numerical 
results for the integrand of the current, ${\rm d}\langle I\rangle/{\rm d}E$
in (\ref{eqn:strom}), in case of a ratio of $A/B=2/1$ are of the same order
of magnitude, as can
be seen in  Figure \ref{fig:dipole}. In the weak-coupling limit it
has been found for the tight-binding scheme that the current vanishes
in linear order when time-reversal parity is present \cite{lehmann,kohler}.
Clearly, there is no weak-coupling limit within the TDSE we solved here and
therefore in our case II, by effectively going to arbitrarily high orders,
we have found a non-vanishing current.
Furthermore, we stress that the results from Floquet scattering theory and 
the converged time-dependent ones shown in Figure \ref{fig:dipole}
coincide within numerical accuracy.

The characteristic jumps in ${\rm d}\langle I\rangle/{\rm d}E(E)$ at
$E=n\omega$ ($n\in\mathbb{N}$) displayed in Figure \ref{fig:dipole} arise
from the fact that at those energies another scattering channel is
opening because $E-n\omega$ becomes bigger than zero. In other words, there
is another Floquet mode with real wavenumber. The sums in
(\ref{eqn:strom}) contain one more term, but that is not the only reason
for the changes in ${\rm d}\langle I\rangle/{\rm d}E$ because then we
would find a step-like change. A characteristic behavior near $E=n\omega$
in the sense of high or discontinuous slopes can already be found for
$|S^{m0}_{\alpha\beta}|^2$ as a function of $E$ (not shown). The reason is 
that the matching conditions are sensible to changes in $E$ when the wavenumber
for a Floquet mode is near zero and changes from imaginary to real 
for increasing $E$. In the insets of Figure \ref{fig:dipole} the
convergence characteristics of the time-dependent results
near $E=\omega$ are displayed and it is shown that increasing
the final time improves the results towards the Floquet result.
Nevertheless, at $E\approx \omega$, the convergence is comparative slow.

\subsection{Time-dependent current}

\begin{figure}[hbtp]
\begin{center}
\includegraphics[clip]{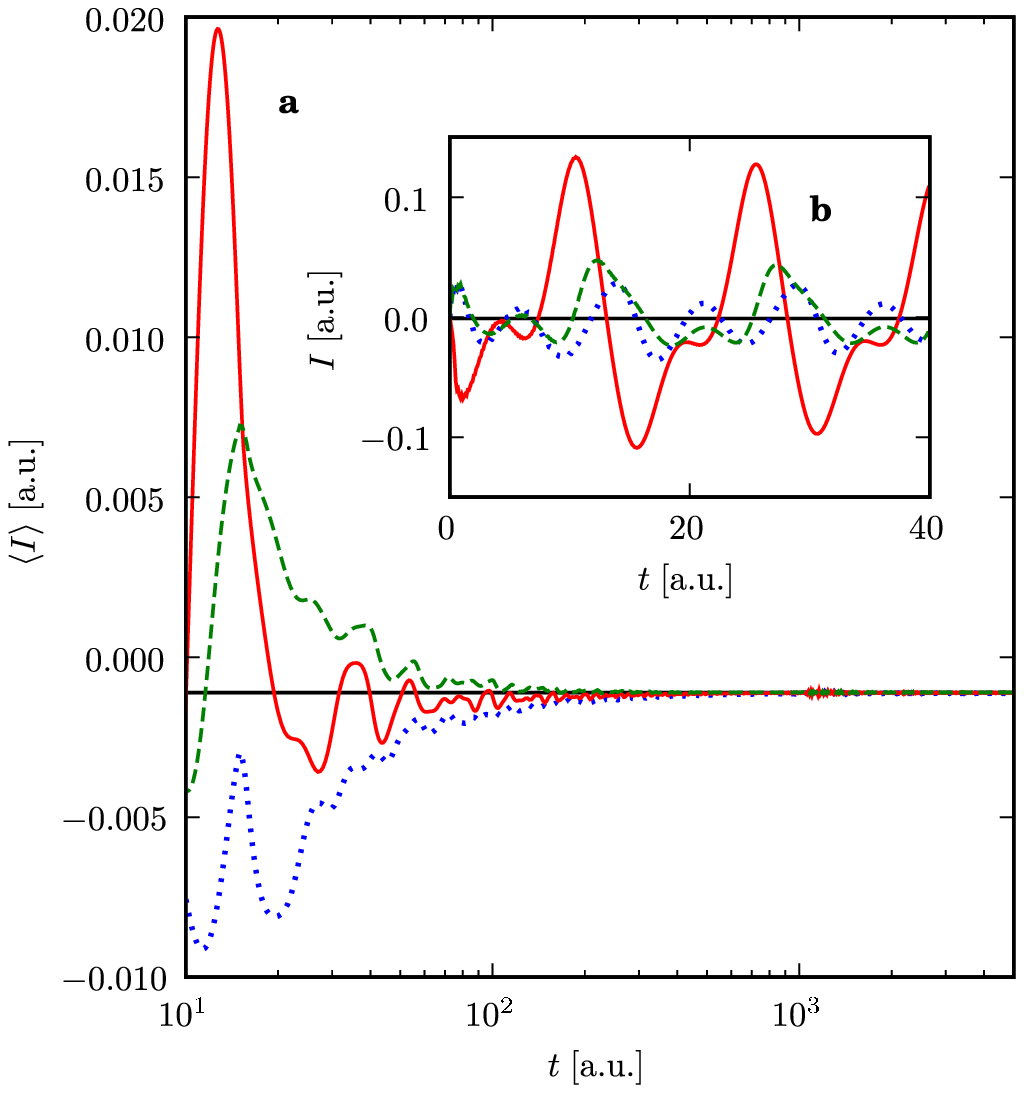}
\caption{(Color online) Time-dependent results for $\langle I\rangle(x,t)$ (a)
and $I(x,t)$ (b) in case I at $x=-d/2$ (blue dotted), $x=0$ (red solid) and
$x=d/2$ (green dashed). Parameters are the same as in Figure \ref{fig:dipole},
$E_{F}=0.3$~a.u. and we have discretized the integration in Eq.
(\ref{eqn:strom_za}) for 100 values of $k$. For comparison the scattering
matrix has been computed with matrices with $11$ Floquet modes and for 2000
energy-values between $0$ and $E_{F}$. Discretized integration according to
(\ref{eqn:strom}) gives $\langle I\rangle=-1.11\times 10^{-3}$~a.u.
(black lines).}
\label{fig:dipole_I_t}
\end{center}
\end{figure}

\begin{figure}[hbtp]
\begin{center}
\includegraphics[clip]{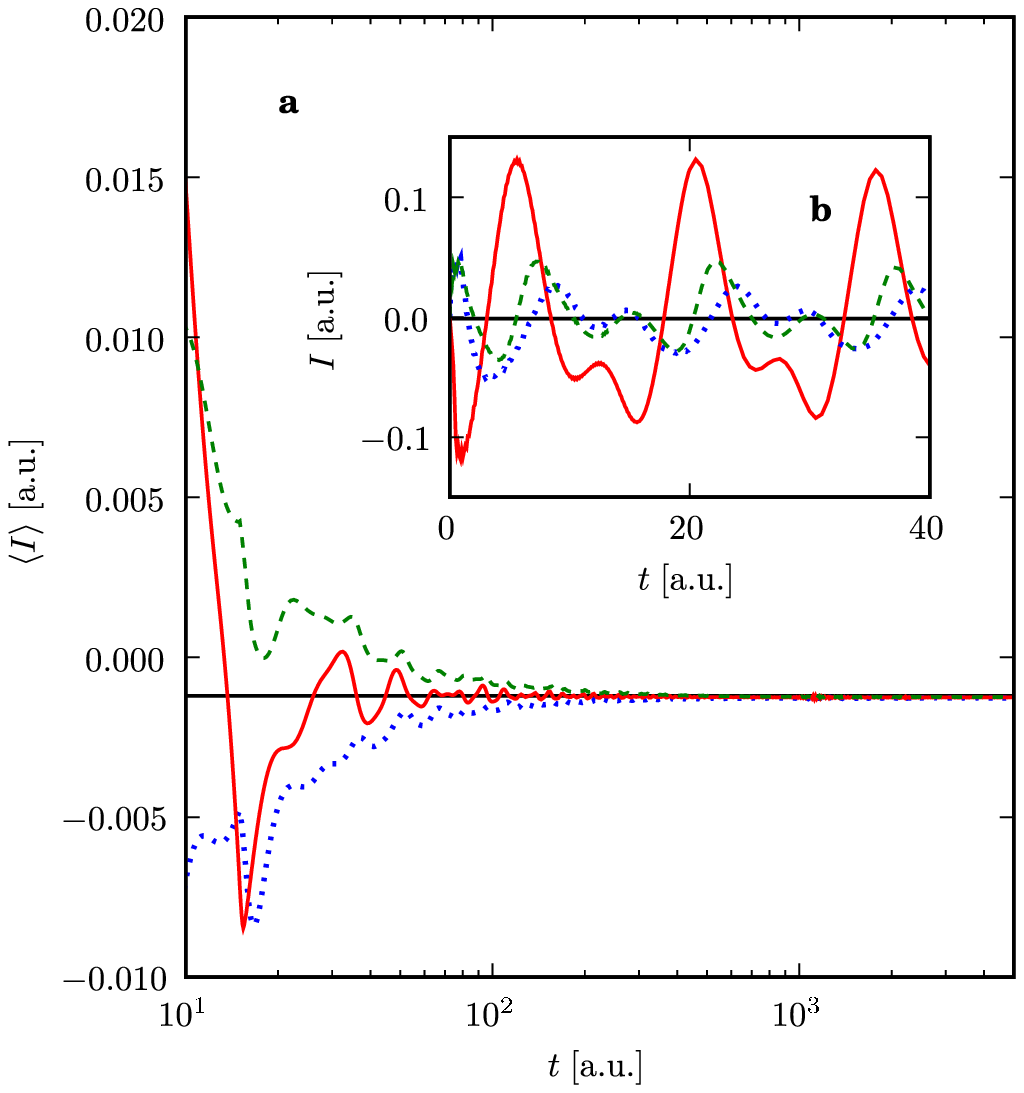}
\caption{(Color online) Time-dependent results for $\langle I\rangle$ (a)
and $I$ (b) in case II at $x=-d/2$ (blue dotted), $x=0$ (red solid) and $x=d/2$
(green dashed). All calculations has been done for the same parameters and in
the same way as for Fig. \ref{fig:dipole_I_t}. The result of Floquet scattering
theory is $\langle I\rangle=-1.21\times 10^{-3}$~a.u. (black lines).}
\label{fig:dipole_I_t2}
\end{center}
\end{figure}

To show the time dependence of the current after switching on the potential
in the cases I and II, $\langle I\rangle(x,t)$ is plotted on a 
logarithmic time scale (Figures \ref{fig:dipole_I_t} and 
\ref{fig:dipole_I_t2}, respectively). As in \cite{stefanucci} the results for
the middle ($x=0$) as well as for the left and right sides ($x=\pm d/2$) of
the central region are shown with different colors (line styles). Although
$I(x,t)$ seems to be nearly time-periodic after only one period of time 
(see insets of Figures \ref{fig:dipole_I_t} and \ref{fig:dipole_I_t2}), there 
are changes in $I(x,t)$ for longer times, which can be seen in the behavior of
$\langle I\rangle(x,t)$ that needs about $10^{3}$~a.u. to coincide
approximately with the result from Floquet scattering theory.

Furthermore, we note that for both potentials,
the amplitude of the current (see insets in Figures \ref{fig:dipole_I_t} and 
\ref{fig:dipole_I_t2}) is much higher at $x=0$ than at $x=\pm d/2$. 
This means, in these dipole fields, there exists periodic forward and backward 
electronic transport which does not arrive at the leads.

\subsection{Optimization of mixing parameter}
\begin{figure}[hbtp]
\begin{center}
\includegraphics[clip]{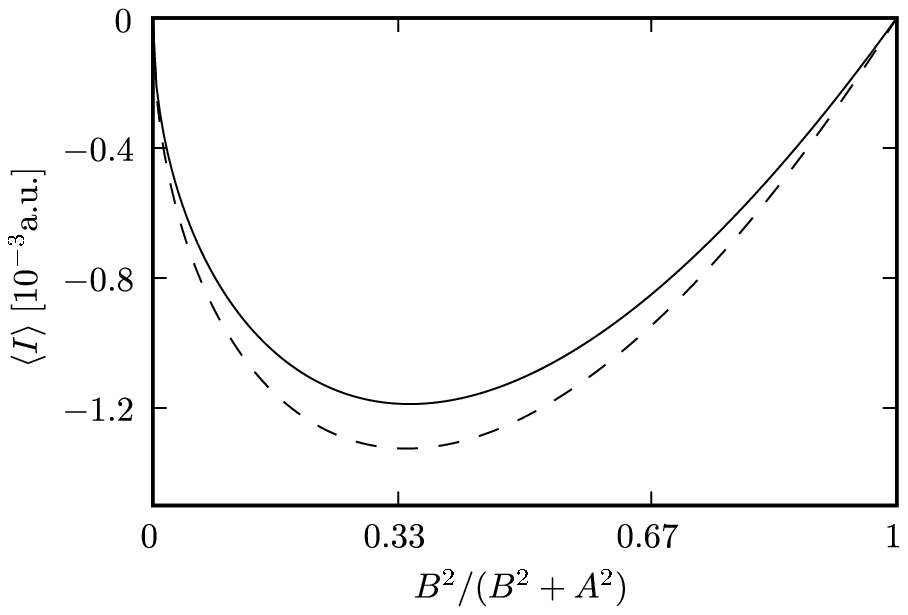}
\caption{$\langle I\rangle$  as a function of $B^2/(B^2+A^2)$  for case I 
(solid) and case II (dashed). The energy integral in (\ref{eqn:strom}) has 
been discretized by using 100 energy values between 0 and 
$E_{F}=0.3~{\rm a.u.}$ Calculations have been performed again with 
$11$ Floquet modes.}
\label{fig:strom}
\end{center}
\end{figure}

So far we had used the ratio $A/B=2/1$ for the relative strength
of first and second harmonic.
In order to find the optimal mixing parameter $B^2/(B^2+A^2)$ 
\footnote{ We have used the ratio $B^2/(A^2+B^2)$ because its limiting
   values are zero and one, allowing for a favorable plotting range 
   of the results.}
we have calculated the average current (\ref{eqn:strom})
for a fixed Fermi energy of $E_{F}=0.3$~a.u. and 
for a fixed value of $A^2+B^2=0.078125$~a.u. using Floquet scattering theory.
We refrain from doing time-dependent calculations 
here as the agreements of the results has been shown above and as the 
time-dependent calculation is numerically more costly.
One finds that to achieve high values of $|\langle I\rangle|$ it is not 
optimal to use a ratio of $A/B=2/1$ (thus $B^2/(A^2+B^2)=0.2$) as done in the
numerical examples before.
The results presented in Fig. \ref{fig:strom} shows a minimum
(i.e. a maximum in the absolute value) in the mixing parameter at
$0.34$ for both cases.

\begin{figure}[hbtp]
\begin{center}
\includegraphics[clip]{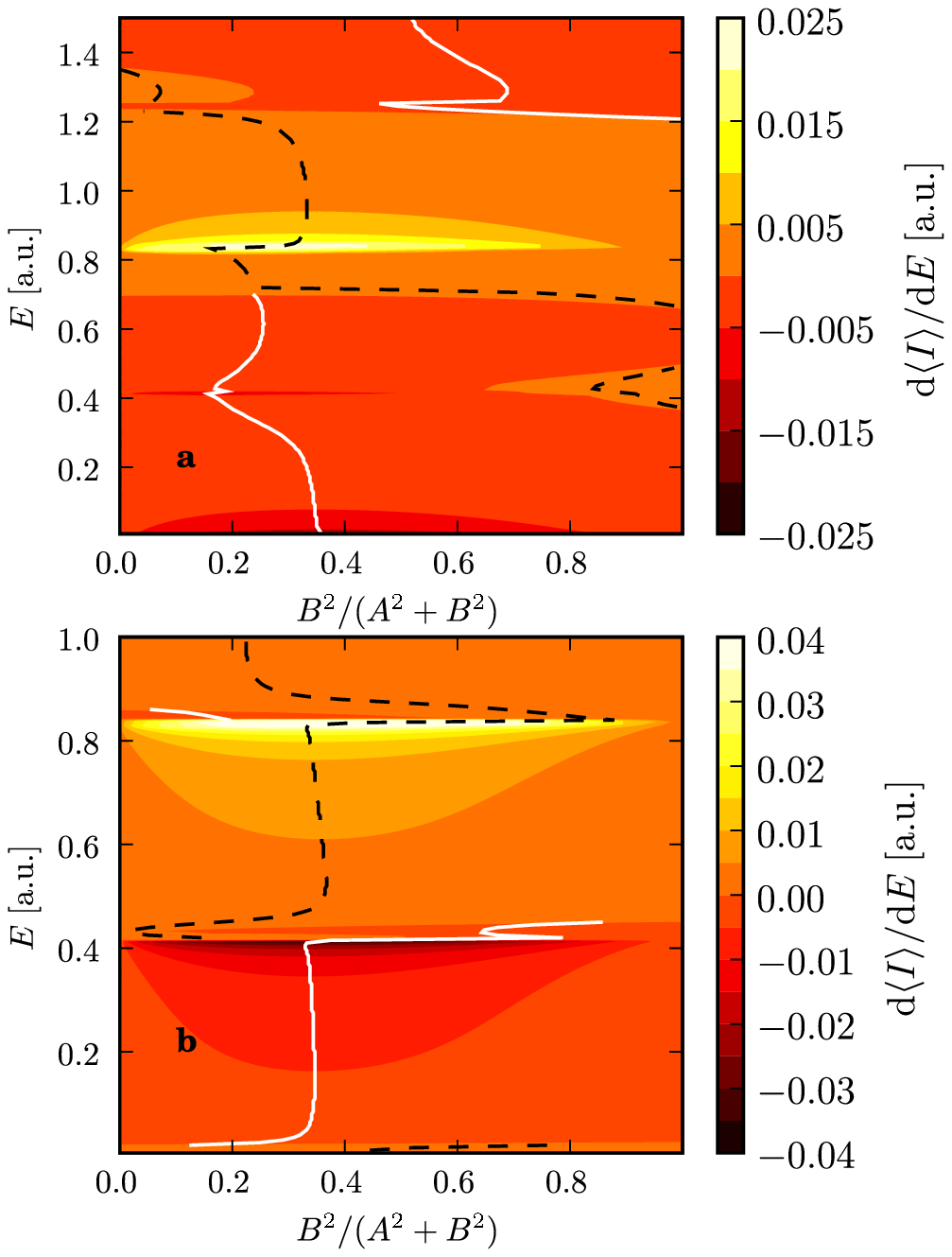}
\caption{(Color online) Contour plot of ${\rm d}\langle I\rangle/{\rm d}E$ as a function
of the incoming energy $E$ and the mixing parameter $B^2/(A^2+B^2)$ with fixed 
$A^2+B^2=0.078125$~a.u. for case I (a) and II (b). The maximum 
(black dashed) and the minimum (white solid lines) of
${\rm d}\langle I\rangle/{\rm d}E$ for fixed $E$ are highlighted. We take once more
Floquet modes with energy between $E-5\omega$ and $E+5\omega$ into account.}
\label{fig:contour}
\end{center}
\end{figure}
In order to elucidate which energy contributes dominantly to
the current, we look at the differential expression 
${\rm d}\langle I\rangle/{\rm d}E$ defined in (\ref{eqn:strom}) 
for different values of the 
incoming energy $E$ and the parameters $A$ and $B$ in (\ref{eqn:dipole1}) and 
(\ref{eqn:dipole2}) again for a fixed value of $A^2+B^2=0.078125$~a.u.
In the limits of either $B=0$ or $A=0$, ${\rm d}\langle I\rangle/{\rm d}E$ is zero 
as generalized parity is valid in these cases. In Figure 
\ref{fig:contour} the results of the calculation are shown in a contour plot. 
The maximum and minimum of ${\rm d}\langle I\rangle/{\rm d}E$ as a function of 
$B^2/(A^2+B^2)$ for fixed $E$ are highlighted.
For several values of $E$ an extremum of 
${\rm d}\langle I\rangle/{\rm d}E$ as a function of $B^2/(A^2+B^2)$ lies around $0.33$ 
thus $B^2/A^2\approx1/2$ especially in case II. On the other hand there are 
no sharp maxima or minima for variable $B$ in contrast to 
${\rm d}\langle I\rangle/{\rm d}E$ as a function of $E$ for a fixed potential. 
Thus $|\langle I\rangle|$ depends more sensitive on the Fermi energy than 
on the mixing parameter.

Finally, it is known from classical \cite{marchesoni} as well as
from tight-binding calculations \cite{goychuk} that typical harmonic mixing
signals for a time-dependent force $F\propto A\cos(\omega t)+B\cos(2\omega t+\phi)$
are in lowest order proportional to $A^2B\cos(\phi)$.
Our cases I and II correspond to $\phi=\pi$ (I) and $\phi=-\pi/2$ (II), respectively.
The proportionality to $A^2B$ is equivalent to a proportionality to
$\sqrt{x}-\sqrt{x^3}$ in the mixing parameter $x:=B^2/(B^2+A^2)$ for constant
$A^2+B^2$, which leads to an extremum at $x=1/3$.
For the situation considered here, the numerical results show that this dependence
on the mixing parameter can be found for a large range of incoming energies
(see Fig. \ref{fig:contour}) and in the integrated current plotted in
Fig. \ref{fig:strom}.
The simple dependence on the phase $\phi$, however, is not observed in general.

\section{Conclusions and Outlook}
\label{sec:concs}
We have investigated a potential with broken (generalized) parity in order to 
generate non-zero net currents in unbiased systems. As expected, in the
long-time limit, the values from time-dependent calculations using
open boundaries agree with the results of the Floquet scattering theory 
obtained from the matching conditions. The characteristic behavior of 
${\rm d}\langle I\rangle/{\rm d}E(E)$ where $E/\omega$ is an integer can be 
understood as the opening of another Floquet channel for the scattering.
Optimization of the mixing parameter $B^2/(A^2+B^2)$ shows that for a large
range of energies a value close to 0.33 leads to a maximum current
amplitude.

In our example the potential is zero in the leads but the applied
methodology can also be used for systems with time-dependent 
potentials in the leads. Floquet scattering theory is numerically 
less costly than time evolution for the case considered here, but it 
is restricted to problems where the time dependence is periodic and the 
analytical solution of the TDSE must be known within certain intervals. 
Moreover a numerical treatment is only possible if a finite number of 
Floquet modes have a substantial contribution to the full solutions 
of the TDSE.

The time-dependent calculation, however, is not restricted to periodic 
problems and it provides information about the time it takes after 
switching on a potential to arrive at the quasi-stationary Floquet results.
This time may be of interest for microelectronics applications. The algorithm 
of time evolution can be used for tight-binding problems as well as for the 
continuous TDSE considered here. For the solution of Kohn-Sham equations 
within TDDFT the algorithm has been used recently by Kurth 
\textit{et al.} \cite{kurth2}.

\section*{Acknowledgement} FG would like to thank Stefan Kurth for
the introduction to the open boundary formalism.
\begin{appendix}

 \section{Floquet S matrix for the harmonic-mixing problem}
\label{sec:Anhang A}

In this appendix we give details of the Floquet approach
to scattering in time-periodic potentials for the harmonic mixing problem
studied herein. 

To start the discussion, we first briefly review the case
of a monochromatic potential of the form 
$V(x,t)=A x \cos(\omega t)$. There the so-called Volkov-solution of 
the  TDSE reads \cite{LR00}
\begin{equation}
 \psi(x,t)=e^{-i f(x,t)}\sum_{n=-\infty}^\infty\left[\tilde a_n e^{iq_ng(x,t)}
+\tilde b_n e^{-iq_ng(x,t)}\right]e^{-in\omega t},
\label{eqn:loessgl}
\end{equation}
wherein the definitions
\begin{eqnarray}
 f(x,t)&=&\epsilon t+\int^t {\rm d}t'~V(x,t')+h(t)=
\epsilon t+\frac{A}{\omega}x\sin(\omega t)-\frac{A^2}{8\omega^3}\sin(2\omega t),
\\
 g(x,t)&=& x+\int^t {\rm d}t'\int^{t'} {\rm d}t''~V(x,t'')/x
        =x-\frac{A}{\omega^2}\cos(\omega t),
\\
 q_n&=&\sqrt{2\left(\epsilon -A^2/(4\omega^2)+n\omega\right)}
\end{eqnarray}
have been used.

The Ansatz (\ref{eqn:loessgl}) is now made for the harmonic-mixing potential 
$V(x,t)=x\left[A\sin(\omega t)+B\cos(2\omega t)\right]$,
with the new definitions
\begin{eqnarray}
 f(x,t)&=&\epsilon t+\int^t {\rm d}t' V(x,t') + h(t)=
\epsilon t+x\left[-\frac{A}{\omega}\cos(\omega t)
+\frac{B}{2\omega}\sin(2\omega t)\right]+h(t),
\\
 g(x,t)&=&x+\int^t {\rm d}t'\int^{t'} {\rm d}t''\frac{V(x,t'')}{x}
=x-\frac{A}{\omega^2}\sin(\omega t)-\frac{B}{4\omega^2}\cos(2\omega t)
\end{eqnarray}
and $q_n=\sqrt{2(\epsilon +C_0+n\omega)}$ with constant $C_0$. For $h(t)$
after some algebra, we get the following ordinary differential equation,
\begin{equation}
 \dot h = \frac{A^2}{4\omega^2}\cos(2\omega t)
          -\frac{B^2}{16\omega^2}\cos(4\omega t)
          -\frac{AB}{4\omega^2}\left[\sin(3\omega t)
          +\sin(\omega t)\right]+\frac{A^2}{4\omega^2}
          +\frac{B^2}{16\omega^2}+C_0.
\end{equation}
The choice $C_0=-A^2/(4\omega^2)-B^2/(16\omega^2)$ warrants that 
$h(t)$ is periodic so that $\epsilon$ is the Floquet energy.
Simple integration leads to
\begin{equation}
 h(t) = \frac{A^2}{8\omega^3}\sin(2\omega t)
        -\frac{B^2}{64\omega^3}\sin(4\omega t)
        +\frac{AB}{12\omega^3}\cos(3\omega t)
        +\frac{AB}{4\omega^3}\cos(\omega t),
\end{equation}
which completes the solution in the central region $C$ in case I of
Sec. \ref{sec:dipole}:
Regarding the potential (\ref{eqn:dipole1}) for $t>0$, the analytical
solution of the TDSE can be written down in each of the regions
$x<-d/2~(L)$, $|x|<d/2~(C)$ and $x>d/2~(R)$,
\begin{eqnarray}
 \psi_L(x,t)&=&e^{-i\epsilon t}\sum_n\left[\frac{a_L^n}{\sqrt{k_n}}e^{ik_nx}
               +\frac{b_L^n}{\sqrt{k_n}}e^{-ik_nx}\right]e^{-in\omega t},\\
\psi_R(x,t)&=&e^{-i\epsilon t}\sum_n\left[\frac{a_R^n}{\sqrt{k_n}}e^{-ik_nx}
               +\frac{b_R^n}{\sqrt{k_n}}e^{ik_nx}\right]e^{-in\omega t},\\
\psi_C(x,t)&=&e^{-i f(x,t)}\sum_n\left[\frac{a^n}{\sqrt{q_n}} e^{iq_ng(x,t)}
               +\frac{b^n}{\sqrt{q_n}}  e^{-iq_ng(x,t)}\right]e^{-in\omega t}
\end{eqnarray}
with $q_n$, $f(x,t)$, $g(x,t)$ as determined before and
$k_n=\sqrt{2 (\epsilon +n\omega)}$.
The matching conditions at $x_L=-d/2$ and $x_R=d/2$ are 
\begin{eqnarray}
\psi_L\left(x_L-0,t\right)&=&\psi_C\left(x_L+0,t\right),
\\
\psi_L'\left(x_L-0,t\right)&=&\psi_C'\left(x_L+0,t\right),
\\
\psi_R\left(x_R+0,t\right)&=&\psi_C\left(x_R-0,t\right),
\\
\psi_R'\left(x_R+0,t\right)&=&\psi_C'\left(x_R-0,t\right).
\end{eqnarray}
Employing the operator $\frac{\omega}{2\pi}\int_0^{\frac{2\pi}{\omega}}{\rm d}t~
e^{is\omega t}$ ($s\in\mathbb{Z}$) on the matching conditions above, i.e.,
make a Fourier transformation, results in
\begin{equation}
\label{eqn:anschluss11}
 \frac{a_L^s}{\sqrt{k_s}}e^{-ik_sd/2}+\frac{b_L^s}{\sqrt{k_s}}e^{ik_sd/2}
 =\sum_n\left[\frac{a^n}{\sqrt{q_n}}e^{-iq_nd/2}C^{L+}_{sn}
              +\frac{b^n}{\sqrt{q_n}}e^{iq_nd/2}C^{L-}_{sn}\right],
\end{equation}
\newline
\begin{equation}
\label{eqn:anschluss12}
\begin{split}
 \sqrt{k_s}\left(a_L^se^{-ik_sd/2}-b_L^se^{ik_sd/2}\right)=\sum_n
         &\left(a^ne^{-iq_nd/2}\left[C^{L+}_{sn}\sqrt{q_n}
              +\frac{D^{L+}_{sn}}{\sqrt{q_n}}\right]\right. \\
         &\left. +b^ne^{iq_nd/2}\left[-C^{L-}_{sn}\sqrt{q_n}
               +\frac{D^{L-}_{sn}}{\sqrt{q_n}}\right]\right),
\end{split}
\end{equation}
\newline
\begin{equation}
\label{eqn:anschluss13}
 \frac{a_R^s}{\sqrt{k_s}}e^{-ik_sd/2}+\frac{b_R^s}{\sqrt{k_s}}e^{ik_sd/2} =
  \sum_n\left[\frac{a^n}{\sqrt{q_n}}e^{iq_nd/2}C^{R+}_{sn}
              +\frac{b^n}{\sqrt{q_n}}e^{-iq_nd/2}C^{R-}_{sn}\right],
\end{equation}
\newline
\begin{equation}
\label{eqn:anschluss14}
\begin{split}
 \sqrt{k_s}\left(-a_R^se^{-ik_sd/2}+b_R^se^{ik_sd/2}\right)=\sum_n
            &\left(a^ne^{iq_nd/2}\left[C^{R+}_{sn}\sqrt{q_n}
               +\frac{D^{R+}_{sn}}{\sqrt{q_n}}\right]\right.\\
            &\left.+b^ne^{-iq_nd/2}\left[-C^{R-}_{sn}\sqrt{q_n}
                +\frac{D^{R-}_{sn}}{\sqrt{q_n}}\right]\right).
\end{split}
\end{equation}
where the definitions
\begin{eqnarray}
C^{\alpha\pm}_{sn}&=&1/T\int_0^T{\rm d}t~
\exp(i\left[(s-n)\omega t -F\left(x_\alpha,t\right)\pm q_nG(t)\right]),
\\
D^{\alpha\pm}_{sn}&=&-1/T\int_0^T{\rm d}t~
F'(t) \exp(i\left[(s-n)\omega t -F\left(x_\alpha,t\right)\pm q_nG(t)\right])
\end{eqnarray}
have been introduced with $\alpha=L,R$ and $F(x,t)=f(x,t)-\epsilon t$, 
$G(t)=g(x,t)-x$, $T=2\pi/\omega$. By introducing the diagonal matrices
$(E)_{km} = \delta_{km}\exp(iq_md/2)$,
$(X)_{km} = \delta_{km}\exp(ik_md/2)$,
$(Q)_{km} = \delta_{km}\sqrt{q_m}$,
$(K)_{km} = \delta_{km}\sqrt{k_m}$,
it is possible to write the matching conditions in matrix form,
\begin{eqnarray}
\label{eqn:an1}
K^{-1}X^{-1}a_L+K^{-1}Xb_L&=&C^{L+}Q^{-1}E^{-1}a+C^{L-}Q^{-1}Eb,\\
\label{eqn:an2}
KX^{-1}a_L-KXb_L&=&\left[C^{L+}QE^{-1}+D^{L+}Q^{-1}E^{-1}\right]a\nonumber\\
& & +\left[-C^{L-}QE+D^{L-}Q^{-1}E\right]b,\\
\label{eqn:an3}
K^{-1}X^{-1}a_R-KXb_R&=&C^{R+}Q^{-1}Ea+C^{R-}Q^{-1}E^{-1}b,\\
\label{eqn:an4}
-KX^{-1}a_R+KXb_R&=&\left[C^{R+}QE+D^{R+}Q^{-1}E\right]a\nonumber\\
& &+\left[-C^{R-}QE^{-1}+D^{R-}Q^{-1}E^{-1}\right]b.
\end{eqnarray}
To get an expression for the Floquet scattering matrix $S$ defined in Eq.
(\ref{eqn:fl-s-matrix}) we need $b_L$ and $b_R$ in dependence of $a_L$ and $a_R$.
By multiplying Eq. (\ref{eqn:an1}) from the left with the matrix $K$ and Eq.
(\ref{eqn:an2}) with $K^{-1}$ it is possible to eliminate the coefficient $b_L$.
For $a_L$ we find the following expression in dependence of $a$ and $b$,
\begin{equation}
\label{eqn:ans1}
\begin{split}
 a_L=&\underbrace{\frac{X}{2}\left[KC^{L+}Q^{-1}
                                   +K^{-1}C^{L+}Q
                                   +K^{-1}D^{L+}Q^{-1}
                              \right]E^{-1}}_{=:B_1}a\\
&+\underbrace{\frac{X}{2}\left[KC^{L-}Q^{-1}
                               -K^{-1}C^{L-}Q
                                +K^{-1}
                         \right]E}_{=:B_2}b.
\end{split}
\end{equation}
In the same manner we can eliminate $b_R$ from Eq. (\ref{eqn:an3}) and Eq. (\ref{eqn:an4})
and get
\begin{equation}
\label{eqn:ans2}
\begin{split}
 a_R=&\underbrace{\frac{X}{2}\left[KC^{R+}Q^{-1}
                                   -K^{-1}C^{R+}Q
                                   -K^{-1}D^{R+}Q^{-1}
                              \right]E}_{=:B_3}a\\
&+\underbrace{\frac{X}{2}\left[KC^{R-}Q^{-1}
                               +K^{-1}C^{R-}Q
                                -K^{-1}D^{R-}Q^{-1}
                          \right]E^{-1}}_{=:B_4}b.
\end{split}
\end{equation}
Eliminating now the coefficients for the incoming waves $a_L$ and $a_R$ from
Eqs. (\ref{eqn:an1}), (\ref{eqn:an2}) or (\ref{eqn:an3}),(\ref{eqn:an4}), respectively,
we find
\begin{equation}
\label{eqn:ans3}
 \begin{split}
  b_L = & \underbrace{\frac{X^{-1}}{2}\left[KC^{L+}Q^{-1}
                                            -K^{-1}C^{L+}Q
                                             -K^{-1}D^{L+}Q^{-1}
                                      \right]E^{-1}}_{=:B_5}a\\
&+\underbrace{\frac{X^{-1}}{2}\left[KC^{L-}Q^{-1}
                                    +K^{-1}C^{L-}Q
                                    -K^{-1}D^{L-}Q^{-1}
                              \right]E}_{=:B_6}b
 \end{split}
\end{equation}
and
\begin{equation}
\label{eqn:ans4}
 \begin{split}
  b_R = & \underbrace{\frac{X^{-1}}{2}\left[KC^{R+}Q^{-1}
                                            +K^{-1}C^{R+}Q
                                            +K^{-1}D^{R+}Q^{-1}
                                      \right]E}_{=:B_7}a\\
&+\underbrace{\frac{X^{-1}}{2}\left[KC^{R-}Q^{-1}
                                    -K^{-1}C^{R-}Q
                                     +K^{-1}D^{R-}Q^{-1}
                              \right]E^{-1}}_{=:B_8}b.
 \end{split}
\end{equation}
Converting Eq. (\ref{eqn:ans1}) leads to $a=B_1^{-1}\left(a_L-B_2b\right)$.
Putting this in Eq. (\ref{eqn:ans2}) results in
$B_3B_1^{-1}\left(a_L-B_2b\right)+B_4b=a_R$.
Converting that to an expression for $b$ yields
\begin{equation}
 \label{eqn:stern}
b=\underbrace{\left[B_4-B_3B_1^{-1}B_2\right]^{-1}}_{=:M_2^{-1}}
   \left(a_R-B_3B_1^{-1}a_L\right).
\end{equation}
One the other hand one gets from Eq. (\ref{eqn:ans2})
$b=B_4^{-1}\left(a_R-B_3a\right)$.
Putting this in Eq. (\ref{eqn:ans1}) then leads to
$\left(B_1-B_2B_4^{-1}B_3\right)a+B_2B_4^{-1}a_R=a_L$
and thus we find
\begin{equation}
 \label{eqn:zweistern}
a=\underbrace{\left[B_1-B_2B_4^{-1}B_3\right] ^{-1}}_{=:M_1^{-1}}
  \left(a_L-B_2B_4^{-1}a_R\right).
\end{equation}
Putting the equations (\ref{eqn:stern}) and (\ref{eqn:zweistern}) in Eqs. (\ref{eqn:ans3})
and (\ref{eqn:ans4}) we directly get
\begin{equation}
 b_L=B_5M_1^{-1}\left(a_L-B_2B_4^{-1}a_R\right)
     +B_6M_2^{-1}\left(a_R-B_3B_1^{-1}a_L\right),
\end{equation}
\begin{equation}
 b_R=B_7M_1^{-1}\left(a_L-B_2B_4^{-1}a_R\right)
     +B_8M_2^{-1}\left(a_R-B_3B_1^{-1}a_L\right).
\end{equation}
We find
\begin{eqnarray}
 S_{LR} & = & -B_5M_1^{-1}B_2B_4^{-1}+B_6M_2^{-1},\\
 S_{RL} & = & B_7M_1^{-1}-B_8M_2^{-1}B_3B_1^{-1}.
\end{eqnarray}
As only $S_{LR}$ and $S_{RL}$ are needed to calculate $\langle I\rangle$ in
Eq. (\ref{eqn:strom}), we have found expressions for the relevant parts of
the Floquet S matrix for the considered system with harmonic mixing.

\end{appendix}

\end{document}